\documentclass[%
 reprint,
 amsmath,amssymb,
 aps,
]{revtex4-2}

\usepackage{graphicx}
\usepackage{dcolumn}
\usepackage{bm}

\begin{document}

\preprint{APS/123-QED}

\title{Speckle Tweezers in Action: Manipulating the Run-and-Tumble of Bacteria}

\author{Ramin Jamali}
\affiliation{Department of Physics, Institute for Advanced Studies in Basic Sciences (IASBS), Zanjan 45137-66731, Iran\vspace{0.1mm}}
\author{Mohammad Hadi Sadri}%
\affiliation{Department of Physics, Institute for Advanced Studies in Basic Sciences (IASBS), Zanjan 45137-66731, Iran}
\author{Ali-Reza Moradi}%
\email{moradika@iasbs.ac.ir}
\affiliation{Department of Physics, Institute for Advanced Studies in Basic Sciences (IASBS), Zanjan 45137-66731, Iran}
\affiliation{School of Nano Science, Institute for Research in Fundamental Sciences (IPM), Tehran 19395-5531, Iran}

\begin{abstract}
This study explores the use of optical speckle tweezers (ST) to manipulate the motility of \textit{Escherichia coli} bacteria. By employing the generated speckle patterns, we demonstrate the ability to control bacterial dynamics through optical forces. Our findings reveal that ST effectively modulates run-and-tumble behavior and alters motility patterns, providing insights into active matter systems. The results establish optical ST as a versatile and non-invasive tool for investigating and controlling the behavior of active particles, offering potential applications in active matter research. 
\end{abstract}

\maketitle
Active matter refers to a class of systems where individual components continuously consume energy and convert it into directed motion. These units, often referred to as ``active particles'', exhibit self-propulsion and collective behaviors that are fundamentally different from passive matter. While passive matter reaches equilibrium when left intact, active matter remains dynamically out-of-equilibrium because its constituent particles convert environmental energy into mechanical motion. This intrinsic non-equilibrium nature opens up a variety of intriguing phenomena \cite{ramaswamy2010mechanics,davis2024active}.
In particular, bacterial baths serve as an exemplary model system for studying active matter, because of their self-propelled behavior and their ability to interact and organize collectively, bacteria, like \textit{Escherichia coli (E.coli)}, are examples of active matter \cite{berg2004coli}. These environments offer a unique platform for studying out-of-equilibrium statistical physics and collective dynamics. Furthermore, one of the most characteristic features of bacterial baths as the mentioned active matter is the run-and-tumble behavior. This motion pattern consists of alternating periods of relatively straight swimming, called ``runs'', and random and sudden reorientations, known as ``tumbles'' \cite{berg2004coli,kurzthaler2024characterization,tailleur2008statistical,paoluzzi2014run}.
This run-and-tumble dynamic is crucial for bacteria to navigate their environment, particularly in response to chemical signals in a process known as chemotaxis \cite{berg1972chemotaxis}.
former methods of controlling active matter often involve applying external fields, like magnetic fields, electric fields, or chemical gradients \cite{bricard2013emergence,martel2006controlled,berg1972chemotaxis,zhang2009artificial,paoluzzi2014run}. However, due to the applications and efficiencies of these methods, these techniques come with limitations, such as the complexity of setup, the lack of fine spatial control, and the possibility of interfering with the biological processes of the bacteria.
Controlling active matter is essential for exploring its dynamics and enabling applications in micro-robotics and biomedical devices \cite{hiratsuka2006microrotary,martel2006controlled,paoluzzi2014run}.
Optical tweezers offer a precise, non-invasive method to control active matter through light-matter interactions, where photon momentum and intensity gradients exert forces on dielectric particles \cite{ashkin1997optical}. Widely used in biology, physics, and materials science \cite{grier2003revolution,padgett2010optical}, their ability to manipulate delicate systems like bacteria without physical contact is particularly advantageous \cite{rey2023light,volpe2006dynamics}.
Optical tweezers excel at manipulating individual particles but struggle with complex, dynamic systems like bacterial suspensions, where collective behaviors and emergent phenomena pose significant challenges to conventional trapping methods \cite{bechinger2016active,grier2003revolution}.
To address these challenges, one of the most promising and recent developments in the applications of optical methods for active matter is the use of speckle tweezers (ST) by employing random optical fields \cite{volpe2014brownian,jamali2021speckle}.
Speckle patterns are generated by the superposition of many phasors in varied methods such as scattering of laser light from a rough surface or mode-mixing in a multimode fiber and transmission of light from complex medium, resulting in a random intensity distribution with bright and dark spots \cite{goodman2007speckle}. This randomness can be harnessed to create multiple optical traps due to the existence of a huge number of intensity gradients. Unlike conventional optical tweezers, which rely on a single focused laser beam, ST provides a flexible and reconfigurable trapping environment. Because of this ability, STs are particularly well-suited for manipulating and trapping large numbers of particles in disordered systems with different physical properties \cite{jamali2024speckle,sadri2024sorting,hanes2012colloids}. The adaptability and scalability of ST make it a versatile tool for external control of the dynamics of active matter systems, for instance, ST can be used to induce controlled fluctuations in the environment of bacteria, allowing for the study of how bacteria respond to complex force fields. Moreover, the speckle pattern is expected to have various applications \cite{jamali2024spargo,sajjadi2024characterization}.
In this paper, we explore the application of optical speckle tweezers to manipulate the run-and-tumble of bacteria. By applying speckle patterns as speckle tweezers, we demonstrate an ST method for controlling active matter. Our results highlight the potential of STs as a powerful tool for studying active matter systems.
To experimentally validate the usage of ST in controlling active matter, we employed a multimode fiber-based ST configuration to manipulate the motility of \textit{E. coli}.
The experimental setup employs a multimode optical fiber to project a 532 nm laser-generated speckle pattern for manipulating the run-and-tumble behavior of \textit{E.coli} (wild-type strain RP437). Mid-log phase bacteria were prepared by culturing, centrifugation, and resuspension in motility buffer to maintain activity. Using optical speckle tweezers, real-time observation and control of bacterial motility were achieved via a microscope-camera system.
The implementation of ST to apply controlled optical forces to \textit{E. coli} bacteria, allowing us to observe the impact on their dynamics and run-and-tumble behavior. When the ST is off (Laser-off), the bacteria exhibit their characteristic random motility, governed primarily by run-and-tumble of bacteria. However, the application of optical forces through speckle patterns alters the motility by imposing external constraints on the bacteria’s motion, reducing the overall motility while maintaining some degree of stochastic movement.
To evaluate the effects of optical forces exerted by the ST, we measured the Mean squared displacement (MSD) of the bacteria under two conditions. The MSD is a critical parameter that quantifies the average displacement of particles over time and provides insights into their diffusion behavior and reflects changes in motility due to external forces.
\begin{figure}[t!]
\centering
\includegraphics[width=0.35\textwidth]{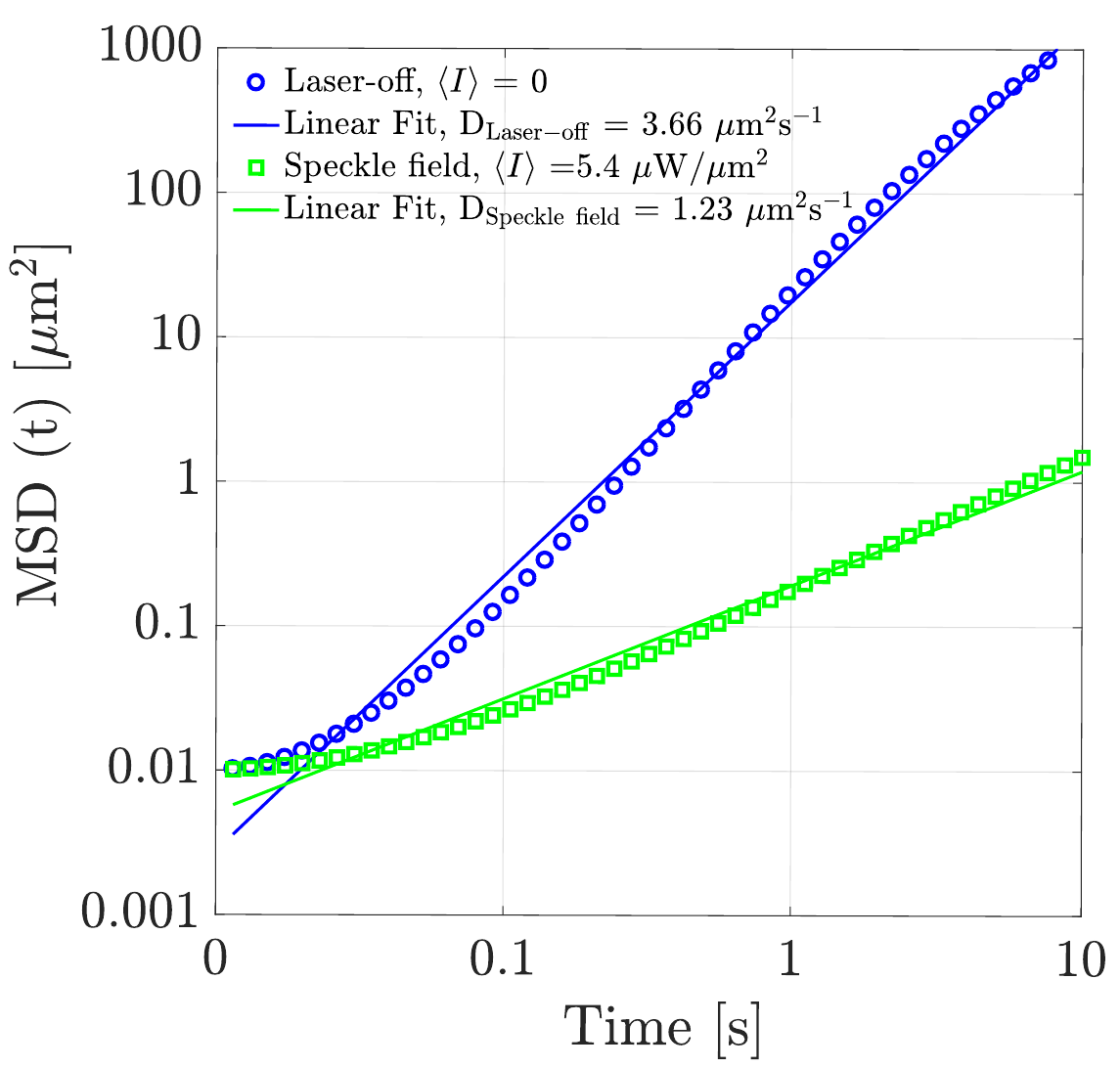}
\caption{Mean squared displacement (MSD) of \textit{E. coli} bacteria under two conditions: laser-off (blue) and ST on (green). The MSD is plotted as a function of time, showing the difference in bacterial motility between the two states. In the absence of optical forces (laser-off), the bacteria exhibit active diffusion, resulting in a steeper MSD curve. When the ST is activated (laser on), the MSD slope decreases significantly, indicating suppression of bacterial motility due to the confining optical forces of the ST.}
\label{fig1}
\end{figure}
Figure \ref{fig1} presents the MSD curves for the two experimental conditions. When the laser is off, the bacteria exhibit typical active motion, characterized by a linear increase in MSD with time, reflecting their unhindered dynamics. However, once the ST is activated, the slope of the MSD curve significantly decreases, indicating that the optical forces applied by the ST have constrained bacterial movement. However, as the ST is turned on, the value of the diffusion coefficient decreases from 3.66 to 1.23 $\mu m^2 s^{-1}$.
The reduction in MSD slope upon switching the ST on confirms that the optical trapping mechanism is successfully limiting the range of bacterial motion. This decrease in motility suggests that the ST imposes a confining force on the bacterial bath, restricting their dynamics and thereby reducing their overall displacement. This effect demonstrates the ability of ST to control bacterial dynamics in real-time, providing a promising and non-invasive tool for studying the behavior of active matter under external control.
\begin{figure}[!t]
\centering
\includegraphics[width=0.35\textwidth]{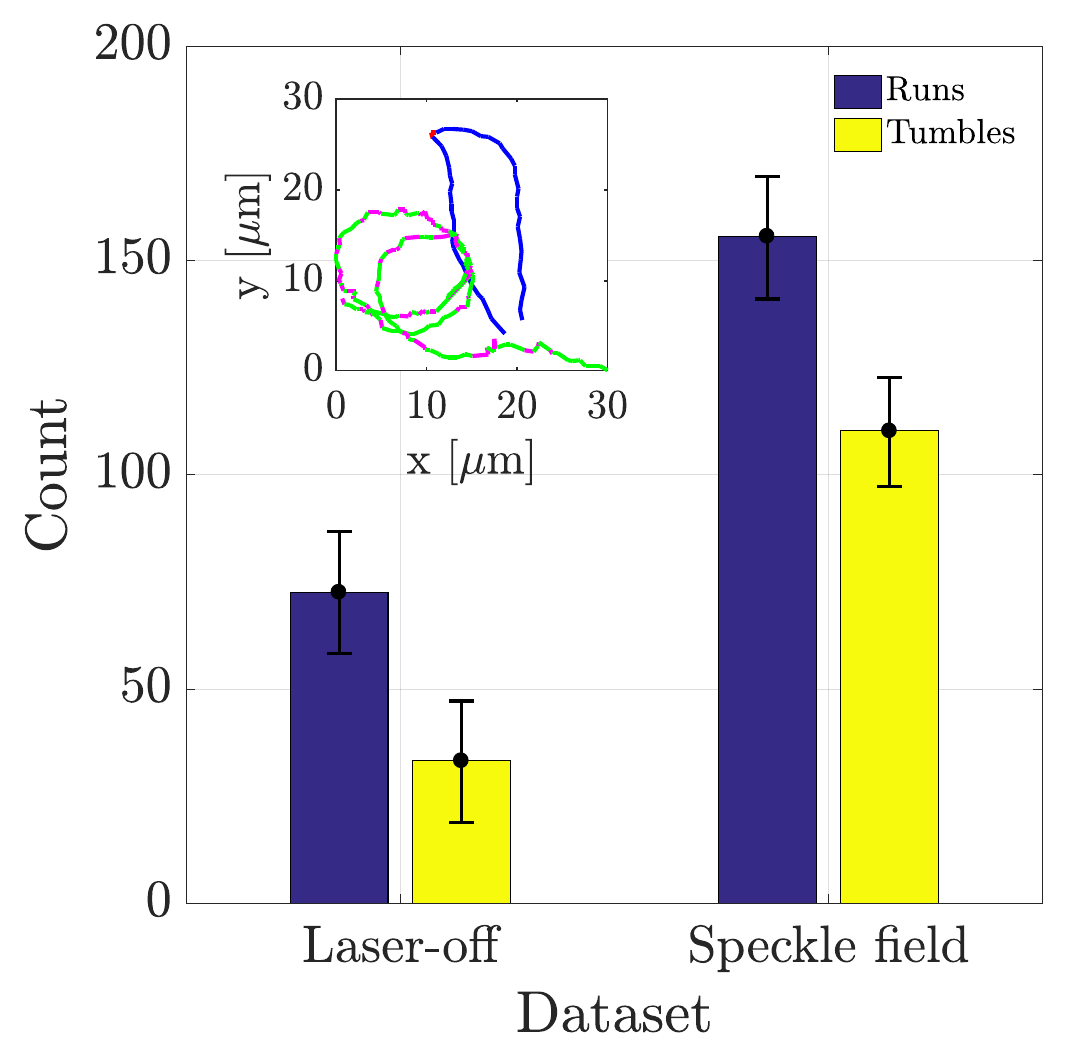}
\caption{The average number of run-and-tumble events for the ensemble of bacteria under two conditions: laser-off and ST on. Both run-and-tumble counts increase under ST, likely due to the confining optical forces altering bacterial motility. Representative trajectories of single bacteria in each condition, showing more frequent direction changes under ST (inset). Trajectories of bacterial motion under speckle field and laser-off conditions.  The blue and red traces represent the run-and-tumble states, respectively, in the absence of a speckle field. The green and magenta traces correspond to the run-and-tumble states under the influence of the speckle field.}
\label{fig2}
\end{figure}
In addition to observing the impact of ST on bacterial motility, we tracked the movement of ensemble \textit{E. coli} bacteria in two conditions: (i) without optical tweezing (laser-off part), where bacteria move freely, and (ii) with the ST activated (ST part). For each bacterium, we calculated the frequency of run-and-tumble steps based on the well-established definitions of these motility modes \cite{berg2004coli}. By averaging the number of run-and-tumble events across the ensemble bacteria in each condition, we observed notable differences in bacterial behavior.
Figure \ref{fig2} presents the mean counts of run-and-tumble events for bacteria in both laser-off and ST conditions. The results show a significant increase in both run-and-tumble frequencies when the ST are activated. This finding suggests that the optical trapping forces imposed by the speckle patterns are influencing bacterial motility by inducing more frequent transitions between the run-and-tumble states. The explanation for this behavior is that the optical trapping forces introduce perturbations into the bacterial bath and related trajectories, causing them to switch between running and tumbling more frequently as they interact with the confining optical forces.
The blue and red traces represent the run-and-tumble states, respectively, in the absence of a speckle field. The green and magenta traces correspond to the run-and-tumble states under the influence of the speckle field.
\begin{figure}[t!]
\centering
\includegraphics[width=0.35\textwidth]{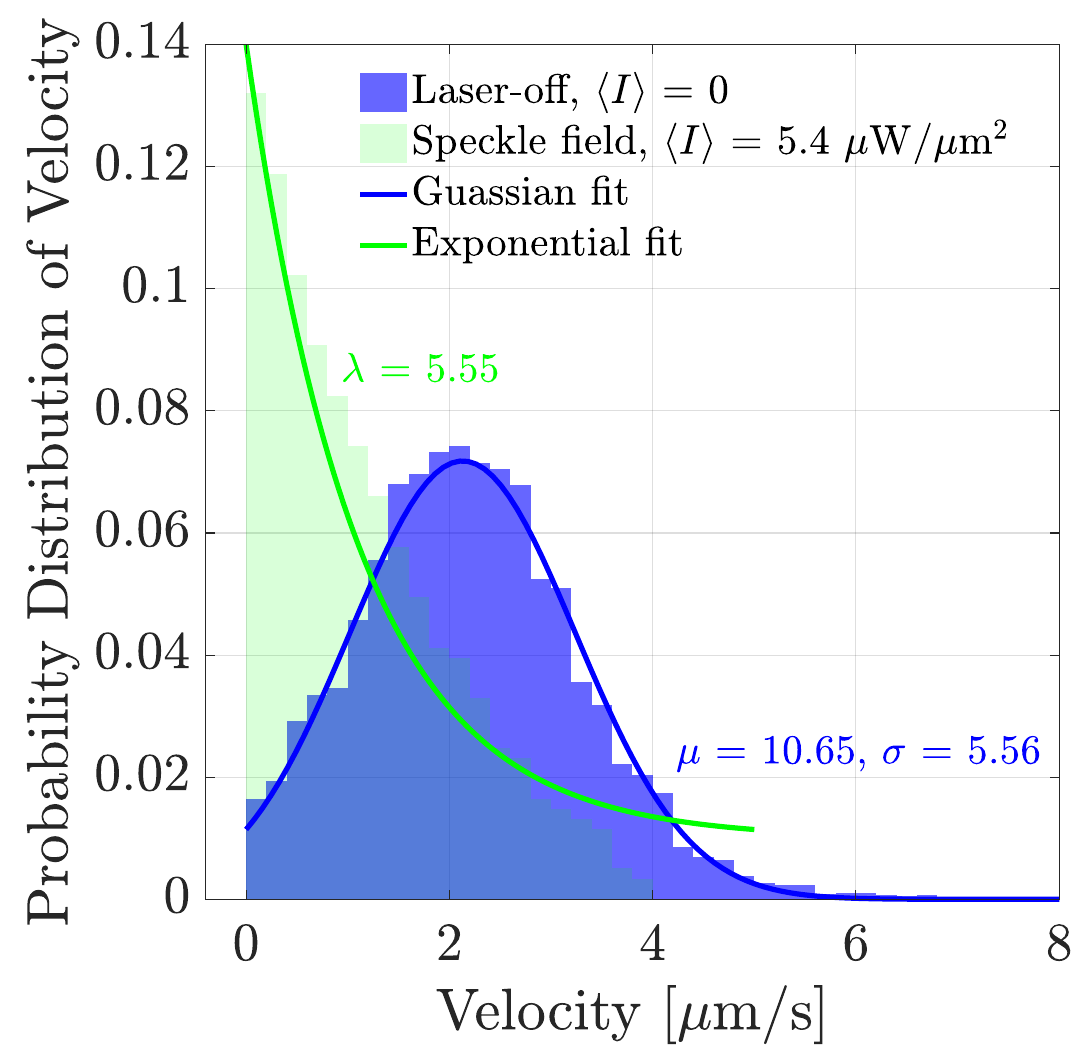}
\caption{Probability distribution of velocity for \textit{E. coli} bacteria in two conditions: laser-off (blue curve) and ST on (green curve). In the laser-off condition, the velocity distribution follows a Gaussian fit, reflecting the natural run-and-tumble behavior of freely moving bacteria. Under ST, the velocity distribution follows a negative exponential fit, indicating that optical forces imposed by the speckle pattern restrict the bacteria’s motion and suppress higher velocities. $\lambda$ is the rate parameter of the exponential distribution. It represents how quickly the probability density decreases. 
$\mu$ is the mean of the Gaussian distribution, which represents the central tendency of the velocity data in the dataset of the laser-off condition.
$\sigma$ is the standard deviation of the Gaussian distribution. It measures the spread or dispersion of the velocity data around the mean ($\mu$).}
\label{fig3}
\end{figure}
To further investigate the impact of ST on bacterial motility, we analyzed the velocity distributions of \textit{E.coli}. Figure \ref{fig3} presents the probability distribution of velocity for particles in two different scenarios, in the laser-off and ST parts. In the laser-off condition, the velocity distribution of the bacteria follows a Gaussian profile, typical for self-propelled particles exhibiting run-and-tumble dynamics. This is consistent with previous studies on bacterial motility \cite{sokolov2015individual,cheong2015rapid}, where the velocity fluctuations stem from the random nature of bacterial motion between running and tumbling phases. The Gaussian fit to the data in this case reflects the absence of external confinement, allowing bacteria to freely explore their surroundings.
In contrast, when the STs are activated, the velocity distribution shifts significantly, following a negative exponential distribution. This change indicates that the optical forces arising from the ST and the intensity gradients within it affect the bacteria’s motion. The negative exponential distribution is a hallmark of systems influenced by trapping forces, such as those exerted by optical tweezers. In this case, the ST confines the bacteria to localized regions, reducing their overall mobility and resulting in a higher probability of lower velocities. The presence of optical forces disrupts the free run-and-tumble of the bacteria, reducing their speed and altering the velocity distribution. The parameter $\lambda$ characterizes the rate at which the probability density decreases. For example, if the velocities of dataset ``ST'' are generally smaller and decay rapidly, $\lambda$ will be high.
The mean ($\mu$) reflects the central tendency of the velocities, while $\sigma$  describes their variability. If the velocities are tightly clustered, $\sigma$ will be small; if they are spread out, $\sigma$  will be larger.
The transition from a Gaussian to an exponential velocity distribution provides evidence that optical forces arising from the ST are restricting bacterial movement. These forces likely reduce the ability of bacteria to perform sustained runs, instead inducing more frequent velocity fluctuations as the bacteria interact with the optical traps. The optical confinement imposed by the speckle field modifies the dynamic behavior of the bacteria, demonstrating that ST can significantly influence the velocity distribution of active particles by exerting localized forces.
\begin{figure*}[t!]
\includegraphics[width=0.99\textwidth]{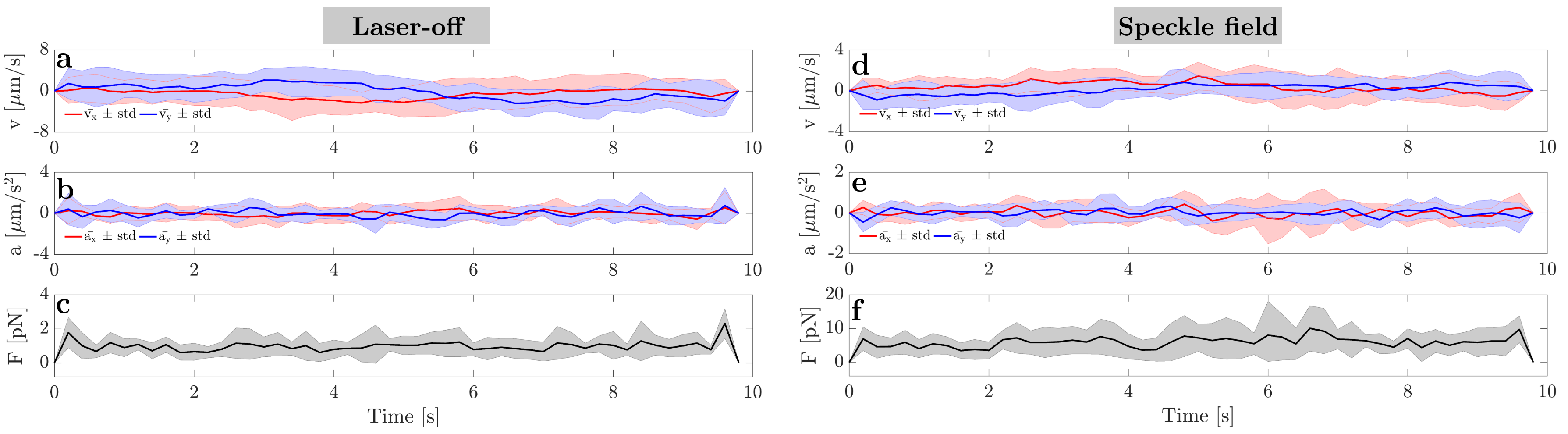}
\caption{Time evolution of velocity, acceleration, and force exerted for \textit{E. coli} bacteria under two conditions: laser-off ((a), (b), and (c)) and ST ((e), (f), and (h)), respectively. In the presence of ST, the optical forces exerted on the bacteria increased, leading to a reduction in both acceleration and velocity. The standard deviations reflect variability in the bacterial responses to the optical forces exerted by the ST.}
\label{fig4}
\end{figure*}
While the velocity distribution highlights how optical forces from the ST influence particle dynamics, a more comprehensive understanding can be obtained by examining the time evolution of key parameters such as force, acceleration, and velocity. To this end, we analyzed these parameters to further quantify the impact of the ST on bacterial motility. These parameters were calculated for all tracked bacteria, and their average values along with standard deviations are presented in Fig. \ref{fig4}. This analysis enables us to assess how the ST alter the dynamics of \textit{E. coli} over time.
When ST is inactive, the force exerted on the bacteria is minimal, and their acceleration and velocity follow typical patterns observed in free-moving bacteria. The velocities tend to be higher, with acceleration fluctuating naturally as bacteria undergo run-and-tumble behavior. However, when the ST are activated, the calculated force exerted on the particles increases significantly. This is a direct result of the optical forces generated by the speckle field, which interact with the bacteria and impose confinement. Consequently, both the acceleration and velocity of the particles decrease because the optical trapping forces restrain the bacteria, leading to reduced motility and thus lower average velocities and accelerations compared to the laser-off condition.
These results are consistent with the hypothesis that the ST introduces  optical forces that actively confine the bacteria, thereby limiting their natural motility. The increase in force, coupled with the reduction in acceleration and velocity, illustrates the ability of the ST to modify bacterial behavior through controlled optical confinement and suppress their movement. 
These findings provide further evidence that ST can be used to manipulate the motility of active matter in a controlled manner, modulating both spatial displacement and dynamic properties such as velocity and acceleration.
\\
In this study, we demonstrated the effectiveness of ST in controlling and manipulating the motility of \textit{E. coli}, a model active matter system. Using a multimode fiber setup to generate speckle patterns, we applied optical forces that altered the bacteria's natural run-and-tumble behavior. This applied method of optical trapping leverages the dynamic and flexible properties of speckle fields, offering scalability and the ability to generate multiple trapping sites.
Our results show a clear impact of ST on bacterial dynamics. Mean squared displacement (MSD) analysis revealed that ST activation significantly reduced bacterial motility, as seen by the decreased slope of the MSD curve. Quantitative tracking showed increased run-and-tumble frequencies under ST, indicating that the optical forces induced more frequent transitions in mobility due to intensity gradients in the traps.
Additionally, the velocity distribution shifted from a Gaussian profile in the laser-off condition to a negative exponential under ST, highlighting the effect of optical confinement. 
Optical forces from the speckle pattern reduced bacterial velocities, disrupting free motility.
Furthermore, time evolution analysis of force, acceleration, and velocity showed that the forces exerted by ST were higher than those in the laser-off condition, resulting in reduced acceleration and velocity. These findings demonstrate that ST provides precise control of active matter systems by modulating their movement through random optical confinement.
In conclusion, ST offers an applied, powerful, and scalable method for manipulating active matter systems like \textit{E. coli}, opening new avenues for studying particle dynamics and collective behavior in complex systems, with promising applications in non-equilibrium phenomena and biophysics.\\
\section*{Data Availability Statement}
The datasets used or analyzed during the current study available from the corresponding author on reasonable request.
\\
\section*{Acknowledgment}
The authors would like to thank Ahmad Farhani Asl for the linguistic editing of the manuscript and Faeze Amarloo for their contributions to the initial stages of the research.

\nocite{*}


\begin{thebibliography}{27}%
\makeatletter
\providecommand \@ifxundefined [1]{%
 \@ifx{#1\undefined}
}%
\providecommand \@ifnum [1]{%
 \ifnum #1\expandafter \@firstoftwo
 \else \expandafter \@secondoftwo
 \fi
}%
\providecommand \@ifx [1]{%
 \ifx #1\expandafter \@firstoftwo
 \else \expandafter \@secondoftwo
 \fi
}%
\providecommand \natexlab [1]{#1}%
\providecommand \enquote  [1]{``#1''}%
\providecommand \bibnamefont  [1]{#1}%
\providecommand \bibfnamefont [1]{#1}%
\providecommand \citenamefont [1]{#1}%
\providecommand \href@noop [0]{\@secondoftwo}%
\providecommand \href [0]{\begingroup \@sanitize@url \@href}%
\providecommand \@href[1]{\@@startlink{#1}\@@href}%
\providecommand \@@href[1]{\endgroup#1\@@endlink}%
\providecommand \@sanitize@url [0]{\catcode `\\12\catcode `\$12\catcode
  `\&12\catcode `\#12\catcode `\^12\catcode `\_12\catcode `\%12\relax}%
\providecommand \@@startlink[1]{}%
\providecommand \@@endlink[0]{}%
\providecommand \url  [0]{\begingroup\@sanitize@url \@url }%
\providecommand \@url [1]{\endgroup\@href {#1}{\urlprefix }}%
\providecommand \urlprefix  [0]{URL }%
\providecommand \Eprint [0]{\href }%
\providecommand \doibase [0]{https://doi.org/}%
\providecommand \selectlanguage [0]{\@gobble}%
\providecommand \bibinfo  [0]{\@secondoftwo}%
\providecommand \bibfield  [0]{\@secondoftwo}%
\providecommand \translation [1]{[#1]}%
\providecommand \BibitemOpen [0]{}%
\providecommand \bibitemStop [0]{}%
\providecommand \bibitemNoStop [0]{.\EOS\space}%
\providecommand \EOS [0]{\spacefactor3000\relax}%
\providecommand \BibitemShut  [1]{\csname bibitem#1\endcsname}%
\let\auto@bib@innerbib\@empty
\bibitem [{\citenamefont {Ramaswamy}(2010)}]{ramaswamy2010mechanics}%
  \BibitemOpen
  \bibfield  {author} {\bibinfo {author} {\bibfnamefont {S.}~\bibnamefont
  {Ramaswamy}},\ }\bibfield  {title} {\bibinfo {title} {The mechanics and
  statistics of active matter},\ }\href@noop {} {\bibfield  {journal} {\bibinfo
   {journal} {Annu. Rev. Condens. Matter Phys.}\ }\textbf {\bibinfo {volume}
  {1}},\ \bibinfo {pages} {323} (\bibinfo {year} {2010})}\BibitemShut {NoStop}%
\bibitem [{\citenamefont {Davis}\ \emph {et~al.}(2024)\citenamefont {Davis},
  \citenamefont {Proesmans},\ and\ \citenamefont {Fodor}}]{davis2024active}%
  \BibitemOpen
  \bibfield  {author} {\bibinfo {author} {\bibfnamefont {L.~K.}\ \bibnamefont
  {Davis}}, \bibinfo {author} {\bibfnamefont {K.}~\bibnamefont {Proesmans}},\
  and\ \bibinfo {author} {\bibfnamefont {{\'E}.}~\bibnamefont {Fodor}},\
  }\bibfield  {title} {\bibinfo {title} {Active matter under control: Insights
  from response theory},\ }\href@noop {} {\bibfield  {journal} {\bibinfo
  {journal} {Physical Review X}\ }\textbf {\bibinfo {volume} {14}},\ \bibinfo
  {pages} {011012} (\bibinfo {year} {2024})}\BibitemShut {NoStop}%
\bibitem [{\citenamefont {Berg}(2004)}]{berg2004coli}%
  \BibitemOpen
  \bibfield  {author} {\bibinfo {author} {\bibfnamefont {H.~C.}\ \bibnamefont
  {Berg}},\ }\href@noop {} {\emph {\bibinfo {title} {E. coli in Motion}}}\
  (\bibinfo  {publisher} {Springer},\ \bibinfo {year} {2004})\BibitemShut
  {NoStop}%
\bibitem [{\citenamefont {Kurzthaler}\ \emph {et~al.}(2024)\citenamefont
  {Kurzthaler}, \citenamefont {Zhao}, \citenamefont {Zhou}, \citenamefont
  {Schwarz-Linek}, \citenamefont {Devailly}, \citenamefont {Arlt},
  \citenamefont {Huang}, \citenamefont {Poon}, \citenamefont {Franosch},
  \citenamefont {Tailleur} \emph {et~al.}}]{kurzthaler2024characterization}%
  \BibitemOpen
  \bibfield  {author} {\bibinfo {author} {\bibfnamefont {C.}~\bibnamefont
  {Kurzthaler}}, \bibinfo {author} {\bibfnamefont {Y.}~\bibnamefont {Zhao}},
  \bibinfo {author} {\bibfnamefont {N.}~\bibnamefont {Zhou}}, \bibinfo {author}
  {\bibfnamefont {J.}~\bibnamefont {Schwarz-Linek}}, \bibinfo {author}
  {\bibfnamefont {C.}~\bibnamefont {Devailly}}, \bibinfo {author}
  {\bibfnamefont {J.}~\bibnamefont {Arlt}}, \bibinfo {author} {\bibfnamefont
  {J.-D.}\ \bibnamefont {Huang}}, \bibinfo {author} {\bibfnamefont {W.~C.}\
  \bibnamefont {Poon}}, \bibinfo {author} {\bibfnamefont {T.}~\bibnamefont
  {Franosch}}, \bibinfo {author} {\bibfnamefont {J.}~\bibnamefont {Tailleur}},
  \emph {et~al.},\ }\bibfield  {title} {\bibinfo {title} {Characterization and
  control of the run-and-tumble dynamics of escherichia coli},\ }\href@noop {}
  {\bibfield  {journal} {\bibinfo  {journal} {Physical Review Letters}\
  }\textbf {\bibinfo {volume} {132}},\ \bibinfo {pages} {038302} (\bibinfo
  {year} {2024})}\BibitemShut {NoStop}%
\bibitem [{\citenamefont {Tailleur}\ and\ \citenamefont
  {Cates}(2008)}]{tailleur2008statistical}%
  \BibitemOpen
  \bibfield  {author} {\bibinfo {author} {\bibfnamefont {J.}~\bibnamefont
  {Tailleur}}\ and\ \bibinfo {author} {\bibfnamefont {M.~E.}\ \bibnamefont
  {Cates}},\ }\bibfield  {title} {\bibinfo {title} {Statistical mechanics of
  interacting run-and-tumble bacteria},\ }\href@noop {} {\bibfield  {journal}
  {\bibinfo  {journal} {Physical review letters}\ }\textbf {\bibinfo {volume}
  {100}},\ \bibinfo {pages} {218103} (\bibinfo {year} {2008})}\BibitemShut
  {NoStop}%
\bibitem [{\citenamefont {Paoluzzi}\ \emph {et~al.}(2014)\citenamefont
  {Paoluzzi}, \citenamefont {Di~Leonardo},\ and\ \citenamefont
  {Angelani}}]{paoluzzi2014run}%
  \BibitemOpen
  \bibfield  {author} {\bibinfo {author} {\bibfnamefont {M.}~\bibnamefont
  {Paoluzzi}}, \bibinfo {author} {\bibfnamefont {R.}~\bibnamefont
  {Di~Leonardo}},\ and\ \bibinfo {author} {\bibfnamefont {L.}~\bibnamefont
  {Angelani}},\ }\bibfield  {title} {\bibinfo {title} {Run-and-tumble particles
  in speckle fields},\ }\href@noop {} {\bibfield  {journal} {\bibinfo
  {journal} {Journal of Physics: Condensed Matter}\ }\textbf {\bibinfo {volume}
  {26}},\ \bibinfo {pages} {375101} (\bibinfo {year} {2014})}\BibitemShut
  {NoStop}%
\bibitem [{\citenamefont {Berg}\ and\ \citenamefont
  {Brown}(1972)}]{berg1972chemotaxis}%
  \BibitemOpen
  \bibfield  {author} {\bibinfo {author} {\bibfnamefont {H.~C.}\ \bibnamefont
  {Berg}}\ and\ \bibinfo {author} {\bibfnamefont {D.~A.}\ \bibnamefont
  {Brown}},\ }\bibfield  {title} {\bibinfo {title} {Chemotaxis in escherichia
  coli analysed by three-dimensional tracking},\ }\href@noop {} {\bibfield
  {journal} {\bibinfo  {journal} {nature}\ }\textbf {\bibinfo {volume} {239}},\
  \bibinfo {pages} {500} (\bibinfo {year} {1972})}\BibitemShut {NoStop}%
\bibitem [{\citenamefont {Bricard}\ \emph {et~al.}(2013)\citenamefont
  {Bricard}, \citenamefont {Caussin}, \citenamefont {Desreumaux}, \citenamefont
  {Dauchot},\ and\ \citenamefont {Bartolo}}]{bricard2013emergence}%
  \BibitemOpen
  \bibfield  {author} {\bibinfo {author} {\bibfnamefont {A.}~\bibnamefont
  {Bricard}}, \bibinfo {author} {\bibfnamefont {J.-B.}\ \bibnamefont
  {Caussin}}, \bibinfo {author} {\bibfnamefont {N.}~\bibnamefont {Desreumaux}},
  \bibinfo {author} {\bibfnamefont {O.}~\bibnamefont {Dauchot}},\ and\ \bibinfo
  {author} {\bibfnamefont {D.}~\bibnamefont {Bartolo}},\ }\bibfield  {title}
  {\bibinfo {title} {Emergence of macroscopic directed motion in populations of
  motile colloids},\ }\href@noop {} {\bibfield  {journal} {\bibinfo  {journal}
  {Nature}\ }\textbf {\bibinfo {volume} {503}},\ \bibinfo {pages} {95}
  (\bibinfo {year} {2013})}\BibitemShut {NoStop}%
\bibitem [{\citenamefont {Martel}\ \emph {et~al.}(2006)\citenamefont {Martel},
  \citenamefont {Tremblay}, \citenamefont {Ngakeng},\ and\ \citenamefont
  {Langlois}}]{martel2006controlled}%
  \BibitemOpen
  \bibfield  {author} {\bibinfo {author} {\bibfnamefont {S.}~\bibnamefont
  {Martel}}, \bibinfo {author} {\bibfnamefont {C.~C.}\ \bibnamefont
  {Tremblay}}, \bibinfo {author} {\bibfnamefont {S.}~\bibnamefont {Ngakeng}},\
  and\ \bibinfo {author} {\bibfnamefont {G.}~\bibnamefont {Langlois}},\
  }\bibfield  {title} {\bibinfo {title} {Controlled manipulation and actuation
  of micro-objects with magnetotactic bacteria},\ }\href@noop {} {\bibfield
  {journal} {\bibinfo  {journal} {Applied Physics Letters}\ }\textbf {\bibinfo
  {volume} {89}} (\bibinfo {year} {2006})}\BibitemShut {NoStop}%
\bibitem [{\citenamefont {Zhang}\ \emph {et~al.}(2009)\citenamefont {Zhang},
  \citenamefont {Abbott}, \citenamefont {Dong}, \citenamefont {Kratochvil},
  \citenamefont {Bell},\ and\ \citenamefont {Nelson}}]{zhang2009artificial}%
  \BibitemOpen
  \bibfield  {author} {\bibinfo {author} {\bibfnamefont {L.}~\bibnamefont
  {Zhang}}, \bibinfo {author} {\bibfnamefont {J.~J.}\ \bibnamefont {Abbott}},
  \bibinfo {author} {\bibfnamefont {L.}~\bibnamefont {Dong}}, \bibinfo {author}
  {\bibfnamefont {B.~E.}\ \bibnamefont {Kratochvil}}, \bibinfo {author}
  {\bibfnamefont {D.}~\bibnamefont {Bell}},\ and\ \bibinfo {author}
  {\bibfnamefont {B.~J.}\ \bibnamefont {Nelson}},\ }\bibfield  {title}
  {\bibinfo {title} {Artificial bacterial flagella: Fabrication and magnetic
  control},\ }\href@noop {} {\bibfield  {journal} {\bibinfo  {journal} {Applied
  Physics Letters}\ }\textbf {\bibinfo {volume} {94}} (\bibinfo {year}
  {2009})}\BibitemShut {NoStop}%
\bibitem [{\citenamefont {Hiratsuka}\ \emph {et~al.}(2006)\citenamefont
  {Hiratsuka}, \citenamefont {Miyata}, \citenamefont {Tada},\ and\
  \citenamefont {Uyeda}}]{hiratsuka2006microrotary}%
  \BibitemOpen
  \bibfield  {author} {\bibinfo {author} {\bibfnamefont {Y.}~\bibnamefont
  {Hiratsuka}}, \bibinfo {author} {\bibfnamefont {M.}~\bibnamefont {Miyata}},
  \bibinfo {author} {\bibfnamefont {T.}~\bibnamefont {Tada}},\ and\ \bibinfo
  {author} {\bibfnamefont {T.~Q.}\ \bibnamefont {Uyeda}},\ }\bibfield  {title}
  {\bibinfo {title} {A microrotary motor powered by bacteria},\ }\href@noop {}
  {\bibfield  {journal} {\bibinfo  {journal} {Proceedings of the National
  Academy of Sciences}\ }\textbf {\bibinfo {volume} {103}},\ \bibinfo {pages}
  {13618} (\bibinfo {year} {2006})}\BibitemShut {NoStop}%
\bibitem [{\citenamefont {Ashkin}(1997)}]{ashkin1997optical}%
  \BibitemOpen
  \bibfield  {author} {\bibinfo {author} {\bibfnamefont {A.}~\bibnamefont
  {Ashkin}},\ }\bibfield  {title} {\bibinfo {title} {Optical trapping and
  manipulation of neutral particles using lasers},\ }\href@noop {} {\bibfield
  {journal} {\bibinfo  {journal} {Proceedings of the National Academy of
  Sciences}\ }\textbf {\bibinfo {volume} {94}},\ \bibinfo {pages} {4853}
  (\bibinfo {year} {1997})}\BibitemShut {NoStop}%
\bibitem [{\citenamefont {Grier}(2003)}]{grier2003revolution}%
  \BibitemOpen
  \bibfield  {author} {\bibinfo {author} {\bibfnamefont {D.~G.}\ \bibnamefont
  {Grier}},\ }\bibfield  {title} {\bibinfo {title} {A revolution in optical
  manipulation},\ }\href@noop {} {\bibfield  {journal} {\bibinfo  {journal}
  {nature}\ }\textbf {\bibinfo {volume} {424}},\ \bibinfo {pages} {810}
  (\bibinfo {year} {2003})}\BibitemShut {NoStop}%
\bibitem [{\citenamefont {Padgett}\ \emph {et~al.}(2010)\citenamefont
  {Padgett}, \citenamefont {Molloy},\ and\ \citenamefont
  {McGloin}}]{padgett2010optical}%
  \BibitemOpen
  \bibfield  {author} {\bibinfo {author} {\bibfnamefont {M.~J.}\ \bibnamefont
  {Padgett}}, \bibinfo {author} {\bibfnamefont {J.}~\bibnamefont {Molloy}},\
  and\ \bibinfo {author} {\bibfnamefont {D.}~\bibnamefont {McGloin}},\
  }\href@noop {} {\emph {\bibinfo {title} {Optical Tweezers: methods and
  applications}}}\ (\bibinfo  {publisher} {CRC press},\ \bibinfo {year}
  {2010})\BibitemShut {NoStop}%
\bibitem [{\citenamefont {Rey}\ \emph {et~al.}(2023)\citenamefont {Rey},
  \citenamefont {Volpe},\ and\ \citenamefont {Volpe}}]{rey2023light}%
  \BibitemOpen
  \bibfield  {author} {\bibinfo {author} {\bibfnamefont {M.}~\bibnamefont
  {Rey}}, \bibinfo {author} {\bibfnamefont {G.}~\bibnamefont {Volpe}},\ and\
  \bibinfo {author} {\bibfnamefont {G.}~\bibnamefont {Volpe}},\ }\bibfield
  {title} {\bibinfo {title} {Light, matter, action: Shining light on active
  matter},\ }\href@noop {} {\bibfield  {journal} {\bibinfo  {journal} {ACS
  photonics}\ }\textbf {\bibinfo {volume} {10}},\ \bibinfo {pages} {1188}
  (\bibinfo {year} {2023})}\BibitemShut {NoStop}%
\bibitem [{\citenamefont {Volpe}\ \emph {et~al.}(2006)\citenamefont {Volpe},
  \citenamefont {Singh},\ and\ \citenamefont {Petrov}}]{volpe2006dynamics}%
  \BibitemOpen
  \bibfield  {author} {\bibinfo {author} {\bibfnamefont {G.}~\bibnamefont
  {Volpe}}, \bibinfo {author} {\bibfnamefont {G.~P.}\ \bibnamefont {Singh}},\
  and\ \bibinfo {author} {\bibfnamefont {D.}~\bibnamefont {Petrov}},\
  }\bibfield  {title} {\bibinfo {title} {Dynamics of a growing cell in an
  optical trap},\ }\href@noop {} {\bibfield  {journal} {\bibinfo  {journal}
  {Applied Physics Letters}\ }\textbf {\bibinfo {volume} {88}} (\bibinfo {year}
  {2006})}\BibitemShut {NoStop}%
\bibitem [{\citenamefont {Bechinger}\ \emph {et~al.}(2016)\citenamefont
  {Bechinger}, \citenamefont {Di~Leonardo}, \citenamefont {L{\"o}wen},
  \citenamefont {Reichhardt}, \citenamefont {Volpe},\ and\ \citenamefont
  {Volpe}}]{bechinger2016active}%
  \BibitemOpen
  \bibfield  {author} {\bibinfo {author} {\bibfnamefont {C.}~\bibnamefont
  {Bechinger}}, \bibinfo {author} {\bibfnamefont {R.}~\bibnamefont
  {Di~Leonardo}}, \bibinfo {author} {\bibfnamefont {H.}~\bibnamefont
  {L{\"o}wen}}, \bibinfo {author} {\bibfnamefont {C.}~\bibnamefont
  {Reichhardt}}, \bibinfo {author} {\bibfnamefont {G.}~\bibnamefont {Volpe}},\
  and\ \bibinfo {author} {\bibfnamefont {G.}~\bibnamefont {Volpe}},\ }\bibfield
   {title} {\bibinfo {title} {Active particles in complex and crowded
  environments},\ }\href@noop {} {\bibfield  {journal} {\bibinfo  {journal}
  {Reviews of modern physics}\ }\textbf {\bibinfo {volume} {88}},\ \bibinfo
  {pages} {045006} (\bibinfo {year} {2016})}\BibitemShut {NoStop}%
\bibitem [{\citenamefont {Volpe}\ \emph {et~al.}(2014)\citenamefont {Volpe},
  \citenamefont {Volpe},\ and\ \citenamefont {Gigan}}]{volpe2014brownian}%
  \BibitemOpen
  \bibfield  {author} {\bibinfo {author} {\bibfnamefont {G.}~\bibnamefont
  {Volpe}}, \bibinfo {author} {\bibfnamefont {G.}~\bibnamefont {Volpe}},\ and\
  \bibinfo {author} {\bibfnamefont {S.}~\bibnamefont {Gigan}},\ }\bibfield
  {title} {\bibinfo {title} {Brownian motion in a speckle light field: tunable
  anomalous diffusion and selective optical manipulation},\ }\href@noop {}
  {\bibfield  {journal} {\bibinfo  {journal} {Scientific reports}\ }\textbf
  {\bibinfo {volume} {4}},\ \bibinfo {pages} {3936} (\bibinfo {year}
  {2014})}\BibitemShut {NoStop}%
\bibitem [{\citenamefont {Jamali}\ \emph {et~al.}(2021)\citenamefont {Jamali},
  \citenamefont {Nazari}, \citenamefont {Ghaffari}, \citenamefont {Velu},\ and\
  \citenamefont {Moradi}}]{jamali2021speckle}%
  \BibitemOpen
  \bibfield  {author} {\bibinfo {author} {\bibfnamefont {R.}~\bibnamefont
  {Jamali}}, \bibinfo {author} {\bibfnamefont {F.}~\bibnamefont {Nazari}},
  \bibinfo {author} {\bibfnamefont {A.}~\bibnamefont {Ghaffari}}, \bibinfo
  {author} {\bibfnamefont {S.~K.}\ \bibnamefont {Velu}},\ and\ \bibinfo
  {author} {\bibfnamefont {A.-R.}\ \bibnamefont {Moradi}},\ }\bibfield  {title}
  {\bibinfo {title} {Speckle tweezers for manipulation of high and low
  refractive index micro-particles and nano-particle loaded vesicles},\
  }\href@noop {} {\bibfield  {journal} {\bibinfo  {journal} {Nanophotonics}\
  }\textbf {\bibinfo {volume} {10}},\ \bibinfo {pages} {2915} (\bibinfo {year}
  {2021})}\BibitemShut {NoStop}%
\bibitem [{\citenamefont {Goodman}(2007)}]{goodman2007speckle}%
  \BibitemOpen
  \bibfield  {author} {\bibinfo {author} {\bibfnamefont {J.~W.}\ \bibnamefont
  {Goodman}},\ }\href@noop {} {\emph {\bibinfo {title} {Speckle phenomena in
  optics: theory and applications}}}\ (\bibinfo  {publisher} {Roberts and
  Company Publishers},\ \bibinfo {year} {2007})\BibitemShut {NoStop}%
\bibitem [{\citenamefont {Jamali}\ \emph
  {et~al.}(2024{\natexlab{a}})\citenamefont {Jamali}, \citenamefont {Velu},\
  and\ \citenamefont {Moradi}}]{jamali2024speckle}%
  \BibitemOpen
  \bibfield  {author} {\bibinfo {author} {\bibfnamefont {R.}~\bibnamefont
  {Jamali}}, \bibinfo {author} {\bibfnamefont {S.~K.}\ \bibnamefont {Velu}},\
  and\ \bibinfo {author} {\bibfnamefont {A.-R.}\ \bibnamefont {Moradi}},\
  }\bibfield  {title} {\bibinfo {title} {Speckle tweezers at fluid-fluid
  interface},\ }\href@noop {} {\bibfield  {journal} {\bibinfo  {journal} {arXiv
  preprint arXiv:2407.07571}\ } (\bibinfo {year}
  {2024}{\natexlab{a}})}\BibitemShut {NoStop}%
\bibitem [{\citenamefont {Sadri}\ \emph {et~al.}(2024)\citenamefont {Sadri},
  \citenamefont {Jamali}, \citenamefont {Khan}, \citenamefont {Rehman},\ and\
  \citenamefont {Moradi}}]{sadri2024sorting}%
  \BibitemOpen
  \bibfield  {author} {\bibinfo {author} {\bibfnamefont {M.~H.}\ \bibnamefont
  {Sadri}}, \bibinfo {author} {\bibfnamefont {R.}~\bibnamefont {Jamali}},
  \bibinfo {author} {\bibfnamefont {A.~J.}\ \bibnamefont {Khan}}, \bibinfo
  {author} {\bibfnamefont {F.}~\bibnamefont {Rehman}},\ and\ \bibinfo {author}
  {\bibfnamefont {A.-R.}\ \bibnamefont {Moradi}},\ }\bibfield  {title}
  {\bibinfo {title} {Sorting of mesoporous silica derivatives by random optical
  fields},\ }\href@noop {} {\bibfield  {journal} {\bibinfo  {journal} {arXiv
  preprint arXiv:2402.14571}\ } (\bibinfo {year} {2024})}\BibitemShut {NoStop}%
\bibitem [{\citenamefont {Hanes}\ \emph {et~al.}(2012)\citenamefont {Hanes},
  \citenamefont {Dalle-Ferrier}, \citenamefont {Schmiedeberg}, \citenamefont
  {Jenkins},\ and\ \citenamefont {Egelhaaf}}]{hanes2012colloids}%
  \BibitemOpen
  \bibfield  {author} {\bibinfo {author} {\bibfnamefont {R.~D.}\ \bibnamefont
  {Hanes}}, \bibinfo {author} {\bibfnamefont {C.}~\bibnamefont
  {Dalle-Ferrier}}, \bibinfo {author} {\bibfnamefont {M.}~\bibnamefont
  {Schmiedeberg}}, \bibinfo {author} {\bibfnamefont {M.~C.}\ \bibnamefont
  {Jenkins}},\ and\ \bibinfo {author} {\bibfnamefont {S.~U.}\ \bibnamefont
  {Egelhaaf}},\ }\bibfield  {title} {\bibinfo {title} {Colloids in one
  dimensional random energy landscapes},\ }\href@noop {} {\bibfield  {journal}
  {\bibinfo  {journal} {Soft Matter}\ }\textbf {\bibinfo {volume} {8}},\
  \bibinfo {pages} {2714} (\bibinfo {year} {2012})}\BibitemShut {NoStop}%
\bibitem [{\citenamefont {Jamali}\ \emph
  {et~al.}(2024{\natexlab{b}})\citenamefont {Jamali}, \citenamefont {Sajjadi},
  \citenamefont {Taherkhani}, \citenamefont {Abbaszadeh},\ and\ \citenamefont
  {Moradi}}]{jamali2024spargo}%
  \BibitemOpen
  \bibfield  {author} {\bibinfo {author} {\bibfnamefont {R.}~\bibnamefont
  {Jamali}}, \bibinfo {author} {\bibfnamefont {M.}~\bibnamefont {Sajjadi}},
  \bibinfo {author} {\bibfnamefont {B.}~\bibnamefont {Taherkhani}}, \bibinfo
  {author} {\bibfnamefont {D.}~\bibnamefont {Abbaszadeh}},\ and\ \bibinfo
  {author} {\bibfnamefont {A.-R.}\ \bibnamefont {Moradi}},\ }\bibfield  {title}
  {\bibinfo {title} {Speckle pattern analysis of pvk: Rgo composite based
  memristor device},\ }\href@noop {} {\bibfield  {journal} {\bibinfo  {journal}
  {Macromolecular Materials and Engineering}\ ,\ \bibinfo {pages} {2400213}}
  (\bibinfo {year} {2024}{\natexlab{b}})}\BibitemShut {NoStop}%
\bibitem [{\citenamefont {Sajjadi}\ \emph {et~al.}(2024)\citenamefont
  {Sajjadi}, \citenamefont {Jamali}, \citenamefont {Kiyani}, \citenamefont
  {Mohamadnia},\ and\ \citenamefont {Moradi}}]{sajjadi2024characterization}%
  \BibitemOpen
  \bibfield  {author} {\bibinfo {author} {\bibfnamefont {M.}~\bibnamefont
  {Sajjadi}}, \bibinfo {author} {\bibfnamefont {R.}~\bibnamefont {Jamali}},
  \bibinfo {author} {\bibfnamefont {T.}~\bibnamefont {Kiyani}}, \bibinfo
  {author} {\bibfnamefont {Z.}~\bibnamefont {Mohamadnia}},\ and\ \bibinfo
  {author} {\bibfnamefont {A.-R.}\ \bibnamefont {Moradi}},\ }\bibfield  {title}
  {\bibinfo {title} {Characterization of schiff base self-healing hydrogels by
  dynamic speckle pattern analysis},\ }\href@noop {} {\bibfield  {journal}
  {\bibinfo  {journal} {Scientific Reports}\ }\textbf {\bibinfo {volume}
  {14}},\ \bibinfo {pages} {27950} (\bibinfo {year} {2024})}\BibitemShut
  {NoStop}%
\bibitem [{\citenamefont {Sokolov}\ \emph {et~al.}(2015)\citenamefont
  {Sokolov}, \citenamefont {Zhou}, \citenamefont {Lavrentovich},\ and\
  \citenamefont {Aranson}}]{sokolov2015individual}%
  \BibitemOpen
  \bibfield  {author} {\bibinfo {author} {\bibfnamefont {A.}~\bibnamefont
  {Sokolov}}, \bibinfo {author} {\bibfnamefont {S.}~\bibnamefont {Zhou}},
  \bibinfo {author} {\bibfnamefont {O.~D.}\ \bibnamefont {Lavrentovich}},\ and\
  \bibinfo {author} {\bibfnamefont {I.~S.}\ \bibnamefont {Aranson}},\
  }\bibfield  {title} {\bibinfo {title} {Individual behavior and pairwise
  interactions between microswimmers in anisotropic liquid},\ }\href@noop {}
  {\bibfield  {journal} {\bibinfo  {journal} {Physical Review E}\ }\textbf
  {\bibinfo {volume} {91}},\ \bibinfo {pages} {013009} (\bibinfo {year}
  {2015})}\BibitemShut {NoStop}%
\bibitem [{\citenamefont {Cheong}\ \emph {et~al.}(2015)\citenamefont {Cheong},
  \citenamefont {Wong}, \citenamefont {Gao}, \citenamefont {Nai}, \citenamefont
  {Cui}, \citenamefont {Park}, \citenamefont {Kenney},\ and\ \citenamefont
  {Lim}}]{cheong2015rapid}%
  \BibitemOpen
  \bibfield  {author} {\bibinfo {author} {\bibfnamefont {F.~C.}\ \bibnamefont
  {Cheong}}, \bibinfo {author} {\bibfnamefont {C.~C.}\ \bibnamefont {Wong}},
  \bibinfo {author} {\bibfnamefont {Y.}~\bibnamefont {Gao}}, \bibinfo {author}
  {\bibfnamefont {M.~H.}\ \bibnamefont {Nai}}, \bibinfo {author} {\bibfnamefont
  {Y.}~\bibnamefont {Cui}}, \bibinfo {author} {\bibfnamefont {S.}~\bibnamefont
  {Park}}, \bibinfo {author} {\bibfnamefont {L.~J.}\ \bibnamefont {Kenney}},\
  and\ \bibinfo {author} {\bibfnamefont {C.~T.}\ \bibnamefont {Lim}},\
  }\bibfield  {title} {\bibinfo {title} {Rapid, high-throughput tracking of
  bacterial motility in 3d via phase-contrast holographic video microscopy},\
  }\href@noop {} {\bibfield  {journal} {\bibinfo  {journal} {Biophysical
  journal}\ }\textbf {\bibinfo {volume} {108}},\ \bibinfo {pages} {1248}
  (\bibinfo {year} {2015})}\BibitemShut {NoStop}%
\end{thebibliography}
\bibliographystyle{unsrt} 

\end{document}